\documentclass[draft,jgrga]{AGUTeX}

\usepackage{lineno}
\linenumbers*[1]

\usepackage[dvips]{graphicx}

\usepackage{amsmath,times,afterpage,color}

 \setkeys{Gin}{draft=false}
\authorrunninghead{CLAUDEPIERRE ET AL.}

\titlerunninghead{ULF PULSATIONS DRIVEN BY DYNAMIC PRESSURE}

\authoraddr{S. G. Claudepierre, The Aerospace Corporation, P.O. Box 92957, MS: M2-260, Los Angeles, CA 90009-2957, USA. (seth.claudepierre@gmail.com)}

\authoraddr{R. E. Denton, Department of Physics and Astronomy, Dartmouth College, Hanover, NH 03755, USA. (richard.e.denton@dartmouth.edu)}

\authoraddr{M. K. Hudson, Department of Physics and Astronomy, Dartmouth College, Hanover, NH 03755, USA. (mary.k.hudson@dartmouth.edu)}

\authoraddr{W. Lotko, Thayer School of Engineering, Dartmouth College, Hanover, NH 03755, USA. (william.lotko@dartmouth.edu)}

\authoraddr{J. G. Lyon, Department of Physics and Astronomy, Dartmouth College, Hanover, NH 03755, USA. (john.g.lyon@dartmouth.edu)}

\begin{document}


\def\eq#1{Equation~(\ref{eq:#1})}
\def\fig#1{Figure~\ref{fig:#1}}
\def\tbl#1{Table~\ref{tbl:#1}}
\def\ch#1{Chapter~\ref{ch:#1}}
\def\sec#1{Section~\ref{sec:#1}}

\def\alfven{Alfv\'{e}n\ }
\def\alfvenic{Alfv\'{e}nic\ }

%
%

\title{Solar wind driving of magnetospheric ULF waves: Field line resonances driven by dynamic pressure fluctuations}


%

%
%

\authors{S. G. Claudepierre, \altaffilmark{1,2} M. K. Hudson, \altaffilmark{1} W. Lotko, \altaffilmark{3} J. G. Lyon, \altaffilmark{1} and R. E. Denton \altaffilmark{1}}

\altaffiltext{1}{Department of Physics and Astronomy, Dartmouth College, Hanover, New Hampshire, USA.}

\altaffiltext{3}{Thayer School of Engineering, Dartmouth College, Hanover, New Hampshire, USA.}

\altaffiltext{2}{Now at The Aerospace Corporation, Los Angeles, CA, USA.}



%
%
%

%
%


\begin{abstract}
Several observational studies suggest that solar wind dynamic pressure fluctuations can drive  magnetospheric ultra-low frequency (ULF) waves on the dayside. To investigate this causal relationship, we present results from Lyon-Fedder-Mobarry (LFM) global, three-dimensional magnetohydrodynamic (MHD) simulations of the solar wind-magnetosphere interaction. These simulations are driven with synthetic solar wind input conditions, where idealized ULF dynamic pressure fluctuations are embedded in the upstream solar wind. In three of the simulations, a monochromatic, sinusoidal ULF oscillation is introduced into the  solar wind dynamic pressure time series.  In the fourth simulation, a continuum of ULF fluctuations over the 0-50 mHz frequency band is introduced into the  solar wind dynamic pressure time series. In this numerical experiment, the idealized solar wind input conditions allow us to study only the effect of a fluctuating solar wind dynamic pressure, while holding all of the other solar wind driving parameters constant.  We show that the monochromatic solar wind dynamic pressure fluctuations drive toroidal mode field line resonances (FLRs) on the dayside, at locations where the upstream driving frequency matches a local field line eigenfrequency.  In addition, we show that the continuum of upstream solar wind dynamic pressure fluctuations drives a continuous spectrum of toroidal mode FLRs on the dayside.  The characteristics of the simulated FLRs agree well with FLR observations, including a phase reversal radially across a peak in wave power, a change in the sense of polarization across the noon meridian,  and a net flux of energy into the ionosphere.  
\end{abstract}

%
%

%

\begin{article}

%

\section{Introduction \label{sec:intro}}

Observations of magnetospheric ultra-low frequency (ULF) pulsations are typically interpreted as the observable signature of magnetohydrodynamic (MHD) waves propagating in the magnetosphere \citep{southwood:83a}.  In this work, `ULF' refers to frequencies in the 0.5-50 mHz range, though we make no distinction between continuous (Pc) and irregular (Pi) pulsations \citep[e.g.][]{jacobs:64a}.   Magnetospheric ULF pulsations can be generated by a variety of mechanisms, which are often classified as either  internal or external mechanisms.  Unstable plasma configurations in the magnetosphere can lead to a local (i.e. internal) generation of ULF pulsations \citep[e.g.][]{hasegawa:69a,hughes:78a}.   In this work, however, we are  concerned with magnetospheric ULF pulsations driven by the solar wind (i.e. externally).  Two examples of external drivers of magnetospheric ULF pulsations are \citep{takahashi:07a}: velocity shear at the magnetopause  (e.g. the Kelvin-Helmholtz instability) and ULF fluctuations in the solar wind dynamic pressure.  Observations of tailward propagating surface waves near the dawn and dusk flank magnetopause are often attributed to the Kelvin-Helmholtz instability  \citep{fairfield:79a,sckopke:81a,hones:81a,couzens:85a}.  Many of the observational features of these magnetopause surface waves agree well with theoretical and numerical studies of the Kelvin-Helmholtz instability \citep{walker:81a,pu:83a,claudepierre:08a}.   However, in this work, we focus on the second external driving mechanism, ULF fluctuations in the solar wind dynamic pressure.

Several observational studies have shown that ULF fluctuations in the solar wind dynamic pressure can drive corresponding ULF fluctuations in the magnetosphere.  For example, \citet{kepko:02a} and \citet{kepko:03a} studied several events where discrete frequency, ULF fluctuations were observed in the solar wind number density. A one-to-one correspondence was observed between the solar wind number density oscillation frequencies and oscillation frequencies in dayside magnetic field measurements from the GOES satellite.  The authors concluded that the fluctuations in the solar wind number density (or dynamic pressure) drove the corresponding fluctuations in the magnetosphere.  \citet{korotova:95a} and \citet{matsuoka:95a} used simultaneous ground based, geosynchronous, and upstream solar wind measurements to show that periodic fluctuations in the solar wind dynamic pressure drove ULF pulsations in the magnetosphere.  \citet{sarafopoulos:95a} reported on a magnetotail lobe oscillation driven by a corresponding fluctuation in the solar wind dynamic pressure. Recent work \citep{viall:09a} has argued that approximately half of the variations observed in magnetospheric ULF waves are driven by solar wind dynamic pressure fluctuations.  While these observational studies have shown that periodic fluctuations in the solar wind dynamic pressure  can drive magnetospheric ULF pulsations in general, other studies have considered the possibility that solar wind dynamic pressure fluctuations can drive field line resonances in the magnetospheric cavity \citep{takahashi:88a,potemra:89a,southwood:90a,stephenson:02a}.

A field line resonance (FLR) is the resonant \alfvenic oscillation of a closed geomagnetic field line, whose foot points lie in the ionosphere.  \citet{dungey:55a} was perhaps the first to suggest that ground based magnetometer observations of discrete frequency fluctuations were the signature of MHD  standing waves on closed geomagnetic field lines.  The discrete frequency oscillations arise from the restriction to an integral number of half-wavelengths along the field line.  This scenario is illustrated schematically in \fig{cartoon} (after \citet{southwood:83a}). The left column shows the standing wave mode structure for the fundamental, second and third harmonics for the electric field oscillation and the right column shows the corresponding mode structure for the magnetic field oscillation.  To simplify the comparisons with our MHD simulations, oscillation amplitude is plotted as a non-negative quantity and, thus, there is no phase information depicted in \fig{cartoon}.  The location of the geomagnetic equatorial plane is indicated by a vertical dashed line in each column and is labeled `eq.'  A perfectly conducting ionosphere is assumed at the field line foot points, which are labeled `iono' in the figure.  At the perfectly conducting (i.e. perfectly reflecting) ionospheric boundaries, the electric field oscillation amplitude goes to zero and the derivative of the magnetic field oscillation amplitude goes to zero \citep{zhu:89a}. This imposes half-wavelength standing waves along the field line.  We note that while the real ionosphere is not a perfect conductor,  magnetospheric FLR mode structure is typically well-approximated in this limit \citep{southwood:83a}.

Ground-based magnetometer observations of ULF pulsations often show a peak in wave power at a particular latitude \citep{samson:71a,samson:72a,walker:79a,ruohoniemi:91a}.  The FLR theories of \citet{chen:74a} and \citet{southwood:74a} predict that this feature is due to the coupling of compressional wave energy into the shear \alfven mode at a spatial location where the local field line eigenfrequency matches the frequency of the compressional source.  The MHD wave equations that describe the coupling between the fast compressional mode and the shear \alfven mode are complicated and have not been completely solved analytically, even in a simple dipole geometry.  We note, however, that some analytic work has been done in nonuniform geometries (including a dipole), leading to approximate solutions \citep[c.f.][]{goedbloed:75a,inhester:86a,chen:89a,mond:90a,wright:94b}.   The equations decouple, however, in the limit of large and small azimuthal mode number, $m$, and describe three modes of oscillation: the poloidal \alfven mode, the toroidal \alfven mode, and the fast mode \citep{radoski:66b,singer:81a}.  In the poloidal mode of oscillation ($m \rightarrow \infty$), dipole magnetic field lines oscillate radially in a fixed meridional plane.  Thus, in cylindrical polar coordinates, the oscillations are in the $v_r$, $B_r$, and $E_{\varphi}$ field components.  In the toroidal mode of oscillation ($m \rightarrow 0$), each $L$-shell oscillates azimuthally and the oscillations are in the $v_{\varphi}$, $B_{\varphi}$ and $E_r$ field components.  In the fast mode of oscillation ($m \rightarrow 0$), the oscillations are in the $v_{r}$, $B_{r}$, $B_{\parallel}$ and $E_{\varphi}$ field components.

The Kelvin-Helmholtz instability at the magnetopause is often cited as the source of compressional energy required for the excitation of FLRs \citep[e.g.][]{chen:74a,southwood:74a}. However, as noted above, several observational studies have suggested that FLRs can be driven by fluctuations in the solar wind dynamic pressure.  For example, \citet{stephenson:02a} reported on HF radar observations of discrete frequency  FLRs, where the same frequencies were observed in solar wind dynamic pressure fluctuations.    \citet{takahashi:88a} reported on an observation of a toroidal mode FLR and the authors suggested that the FLR was driven by a series of solar wind pressure disturbances.   \citet{potemra:89a} considered simultaneous observations of the solar wind, measured by the IMP 8 satellite,  and the magnetosphere, measured by ground based magnetometers and the AMPTE satellite.  The authors concluded that a monochromatic fluctuation in the solar wind number density launched a tailward traveling wave train in the magnetosphere that excited local FLRs.  In addition to these observational studies, the theoretical work of \citet{southwood:90a} studied the magnetospheric response to solar wind dynamic pressure changes.  They found both FLRs and a global, compressional eigenoscillation of the entire magnetospheric cavity, known as the `cavity mode,' in the response.  These studies all suggest that solar wind dynamic pressure fluctuations can provide the compressional source of energy required for the excitation of FLRs.

In this work, we use global MHD simulations to study the role that solar wind dynamic pressure fluctuations play in the generation of magnetospheric ULF pulsations.  We show that monochromatic ULF fluctuations in the solar wind dynamic pressure can drive toroidal mode field line resonances in the simulation magnetosphere.  In addition, we show that fluctuations over a continuum of ULF frequencies in the  solar wind dynamic pressure can drive a continuous spectrum of toroidal mode FLRs.   The remainder of this paper is structured as follows:  in \sec{method}, we discuss the Lyon-Fedder-Mobarry global MHD simulation, which is the primary tool used in this work and, in \sec{simres}, we present the simulation results.  \sec{discuss} provides a detailed  discussion of the simulated FLRs, including a   comparison with some general characteristics of FLR observations.  Finally, in \sec{conclu}, we provide a summary of the results and offer some concluding remarks.

\section{Methodology \label{sec:method}}

The Lyon-Fedder-Mobarry (LFM) code solves the single fluid, ideal magnetohydrodynamic (MHD) equations on a three-dimensional grid to simulate the interaction between the solar wind, magnetosphere, and ionosphere.  We discuss the features of the LFM simulation that are the most pertinent to this study; a  more detailed code description, including a discussion of the accuracy of the solution, can be found in \citet{lyon:04a}.

The LFM  computational grid is a non-Cartesian, distorted spherical mesh, which, in this study, contains 106x48x64 grid cells.  This translates to a spatial resolution of roughly 0.25 $R_E$ in the inner magnetosphere ($r < 6.6$ $R_E$), though the spacing is not uniform.  The outer boundary condition in the LFM simulation is the solar wind, which is input into the simulation at the upstream boundary at $x= \text{30 } R_E$ (GSM coordinates are used throughout this study).  In fact, the solar wind boundary conditions are only imposed at the upstream boundary; supersonic outflow conditions are used along the rear boundaries and ballistic propagation on the side boundaries.  The solar wind input conditions propagate as 2D planar fronts, from the upstream boundary to the magnetosphere.  The input conditions can be either real solar wind data taken from satellite measurements (e.g. ACE or WIND) or idealized configurations. The LFM simulations presented in this study are driven by idealized solar wind input conditions with the solar wind variations traveling in the negative $x$-direction as $yz$-planar fronts.  The dipole tilt of the geomagnetic field is neglected so the SM and GSM coordinate systems are the same in this study.  The propagation of the solar wind as planar fronts and the zero dipole tilt angle both affect the field line resonance mode structure in the simulation magnetosphere.

For the inner boundary condition, the magnetospheric portion of the LFM simulation couples to a 2D electrostatic model of the ionosphere \citep{fedder:95a}.  Within this model, the electric potential is obtained from Poisson's equation.  The potential is then mapped along dipole field lines back to the inner boundary of the LFM simulation, which is located at a geocentric distance of roughly 2 $R_E$.  At the LFM inner boundary, the mapped ionospheric potential is used to calculate the electric field, which serves as part of the boundary condition for the magnetospheric portion of the LFM.  The ionospheric conductance, needed for the potential solver, is specified by an empirical extreme ultraviolet (EUV) conductance model and by a contribution from particle precipitation.  The EUV conductance model, similar to the one used in the AMIE model \citep{richmond:92a}, is parametrized by the solar EUV flux.  Within the LFM code, the solar 10.7 cm flux is used to regulate the solar EUV flux and is specified as an input to the ionospheric model.  In the idealized LFM simulations presented in this work,  the choice for the 10.7 cm flux is somewhat arbitrary and we choose a value of $\text{100} \times \text{10}^\text{-22}$ J/m$^2$ to represent a `typical' value.  Within the ionospheric model, the EUV conductance is also modified by electron precipitation from the magnetosphere, with properties derived from the MHD variables at the LFM inner boundary.  This entire computation results in a two-way coupling between the ionospheric model and the magnetospheric portion of the code, as described in greater detail by \citet{wiltberger:09a}.

\subsection{Simulation Details \label{sec:deets}}

We present results from four LFM simulations: three driven by monochromatic ULF fluctuations in the solar wind dynamic pressure and one driven by a continuum of ULF frequencies in the upstream solar wind dynamic pressure fluctuations.  Solar wind dynamic pressure (henceforth, $p_{dyn}$) is not an explicit input  in the LFM simulation and we choose to introduce the $p_{dyn}$ fluctuations via the upstream number density, as opposed to the velocity ($p_{dyn} \propto n v^2$).  Solar wind observations typically show that $p_{dyn}$ variations are carried by the solar wind number density rather than the velocity \citep[e.g.][]{kepko:03a,han:07a}. For the three monochromatic simulations, we impose a number density time series, $n(t)$, at the LFM upstream boundary at $x=\text{30}$ $R_E$ of the form:  
\begin{equation*}
        n(t)=n_{0}+\delta n \sin (\omega t)
\label{eq:mono}
\end{equation*}
Here, $\omega=2 \pi f$ is the monochromatic driving frequency, $n_{0}$ = 5 particles/cm$^3$ is the background number density, and $\delta n$ is amplitude of the perturbation.  The three monochromatic driving frequencies chosen for analysis in this study are 10, 15, and 25 mHz. In the 10 mHz and 15 mHz simulations, $\delta n= \text{1}$ (20$\%$ oscillation amplitude) while in the 25 mHz simulation, $\delta n= \text{2}$ (40$\%$ oscillation amplitude).  The larger oscillation amplitude for the input time series in the 25 mHz simulation is used to combat the effects of a numerical attenuation/filtering of higher frequency components in the LFM simulation.  This complication is discussed in greater detail in the next subsection.

In the fourth simulation, we impose a continuum of ULF frequencies in the input number density time series:
\begin{equation}
        n(t)=n_{0}+0.05 \sum_{j} \sin (\omega_j t + \phi_j)
\label{eq:contin}
\end{equation}
Here, we define $\omega_j = 2 \pi f_{j} = 2 \pi j/10000$ and carry out the summation from $j = 0,1,\cdots, 500$.  This choice defines an input time series with a spectrum that contains a superposition of discrete frequency fluctuations in the 0 to 50 mHz band and a 0.1 mHz spacing between spectral components.  We also add a random phase, $\phi_j$, to each spectral component.  The value of 0.05 in the above equation is chosen so that the root-mean square ($RMS$) amplitude of the   continuum input number density time series is roughly equal to that of the  monochromatic input number density time series (with 20$\%$ oscillation amplitudes).  In addition, in all four simulations, we introduce an appropriate out of phase oscillation in the input sound speed time series, so as to hold the thermal pressure constant in the upstream solar wind ($p_{th} \propto nC_{s}^{2}$). The background sound speed upon which the out of phase oscillation is imposed is 40 km/s.  The remaining idealized solar wind input parameters  are the same in all four simulations and held constant during the four hour intervals selected for analysis:  {\bf B} = (0,0,5) nT and {\bf v} = (-600,0,0) km/s.  We will refer to the three  monochromatic simulations as the ``10 mHz,'' ``15 mHz,'' and ``25 mHz simulations'' and to the fourth simulation as the ``continuum simulation.''  The nature of the driving in the continuum simulation is chosen to examine the magnetospheric response to fluctuations over a range of ULF frequencies.  Although the input spectrum is not a true continuum, it represents a quasi-broadband driving spectrum derived from quasi-random fluctuations and is sufficient for the purposes of this study.

In the four simulations presented in this study, an interval of preconditioning is run prior to the four hour  periods selected for analysis.  The simulations begin with one hour of constant southward interplanetary magnetic field (IMF) of $-$5 nT, followed by one hour of constant northward IMF ($B_z = +5$ nT), followed by two hours of constant southward IMF ($B_z = -5$ nT), followed by one hour of constant northward IMF ($B_z = +5$ nT).  During this initial five hour preconditioning phase, all of the remaining solar wind parameters are held constant, including the number density and sound speed ($n(t) = 5$ particles/cm$^3$ and $C_{s}(t) = 40$ km/s).  The oscillations in $n(t)$ and $C_{s}(t)$ described above begin at the end of the five hour preconditioning period.  In our analyses, we do not wish to include the initial response of the magnetosphere to the `turning-on' of the solar wind fluctuations.  Thus, we examine the four hours of simulation time after the first hour of upstream fluctuations.  In summary, the entire duration of the simulations is 10 hours: hours 0-5 are the preconditioning phase, hours 5-10 are the upstream driving phase and the periods selected for analysis are from hours 6-10.

\subsection{Simulation Caveats \label{sec:caveats}}

\begin{figure}
\begin{center}
 \includegraphics[scale=0.35]{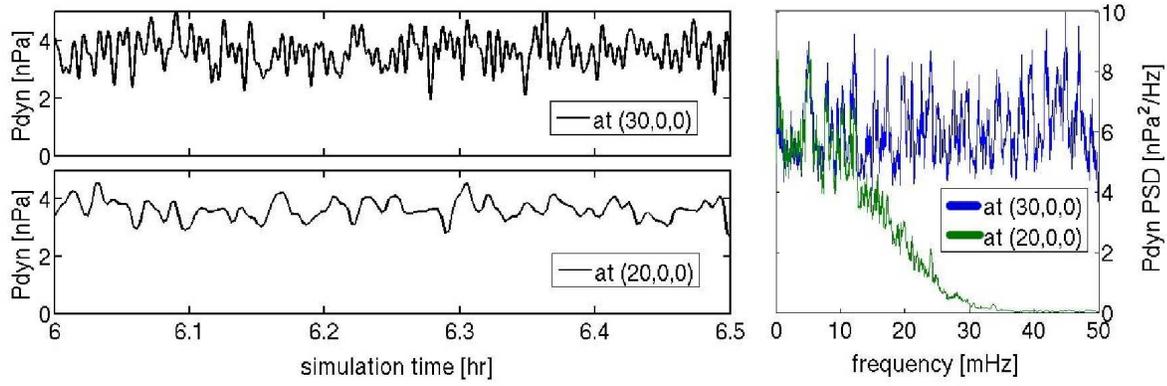}
\end{center}
 \caption{ Solar wind dynamic pressure driving in the continuum simulation. The two time series on the left show the dynamic pressure  input into the simulation at the upstream boundary (at {\bf x} = (30,0,0) $R_E$, top panel) and the resultant dynamic pressure in the upstream solar wind in the simulation 10 $R_E$ downstream (at {\bf x} = (20,0,0) $R_E$, bottom panel).  Note that the higher frequency fluctuations in the input time series are filtered out by the time the fluctuations reach {\bf x} = (20,0,0) $R_E$ in the simulation.  The blue and green traces in the right panel are the power spectral densities of the two time series shown on the left, illustrating the high frequency filtering.  The green trace is the spectral profile that drives the magnetosphere in the continuum simulation.   }
\label{fig:filter}
\end{figure}

Several numerical techniques used in the LFM model influence the simulated wave dynamics considered here.  These techniques are briefly discussed to facilitate later interpretations.  As alluded to above, the oscillation amplitudes of the higher frequency $p_{dyn}$ fluctuations imposed at the upstream boundary are attenuated in the simulations.  We illustrate this phenomenon in the  continuum simulation, though a similar effect is encountered in the monochromatic simulations.  The two panels on the left in \fig{filter} show time series of the solar wind $p_{dyn}$ imposed at the upstream boundary at (30,0,0) $R_E$ (top panel) and at (20,0,0) $R_E$ in the upstream solar wind in the simulation (bottom panel).  Solar wind $p_{dyn}$ is shown on the vertical scales in both panels and ranges from 0 to 5 nPa.  Simulation time is shown in hours on the horizontal scale in both panels and ranges from 6.0 to 6.5 hrs (decimal time).   The time series plotted in the top panel is that derived from \eq{contin} wherein $p_{dyn} = m_{i} n(t) v_{sw}^2$.  Comparison of the two time series shows that the higher frequency spectral components imposed at the upstream boundary are attenuated downstream from the boundary.  In the right panel in \fig{filter}, the power spectral densities ($PSD$) of the full four hour $p_{dyn}$ time series are shown in blue (at (30,0,0) $R_E$) and in green (at (20,0,0) $R_E$).   The blue trace shows the relatively uniform distribution of wave power across the entire 0 to 50 mHz spectral band, as expected from the Fourier transform of \eq{contin}.  Comparison of the two traces shows that the spectral profile input at the upstream boundary is significantly altered by the time the fluctuations reach (20,0,0) $R_E$ in the upstream solar wind.  For frequencies in the 0 to 10 mHz range, little to no modification of the input spectrum is evident.  For frequencies in the 10 to 35 mHz range, the oscillation amplitudes in the input spectrum are significantly attenuated.  For frequencies in the 35 to 50 mHz range, the oscillations imposed at the upstream boundary are almost completely suppressed (attenuated) in the downstream region.

The attenuation of the higher frequency components imposed at the upstream boundary results from the numerics in the LFM simulation. The LFM uses nonlinear, total variation diminishing (TVD) switches to allow shock capturing. These switches also cause dissipation as the wavelength of a wave approaches the grid scale. Linear dispersion studies with the LFM show that this effect becomes significant when a full wavelength is resolved by about 6 grid cells or less. For the grid used in these studies, this condition corresponds to $\approx$ 3 $R_E$ in the upstream solar wind and a temporal frequency of $\approx$ 30 mHz. The estimate from \fig{filter} shows that the damping rate at 30 mHz is approximately half of the real part of $\omega$. This attenuation is somewhat higher than what is seen from the dispersion of linear waves. The discrepancy is not large and can be explained by two factors. The waves introduced are already nonlinear, will tend to self-steepen, and cause the TVD switches to turn on. In addition, the upwind boundary condition does not attempt to create a solution accurate to the underlying eighth-order accuracy of the LFM transport. The damping rate of the LFM switches depends on the accuracy of the solution outside of discontinuous regions. The more accurate the solution, the lower the effective damping rate is.  Nonetheless, the spectral profile that drives the magnetosphere (\fig{filter}, green trace), contains a continuum of ULF frequencies in the $\approx$ 0-25 mHz range and is sufficient for the purposes of this study.   We emphasize that the magnetosphere is driven by a continuum of ULF frequencies in the upstream $p_{dyn}$ fluctuations.  Later, we will see that the magnetospheric response selects particular frequencies from this continuum of upstream fluctuations.

The attenuation of the amplitudes in the input time series in the continuum simulation results in $RMS$ upstream driving at (20,0,0) $R_E$ on the order of 13$\%$, reduced from the roughly 20$\%$ value imposed at the upstream boundary (in the $RMS$ sense described above).  As noted above, the filtering/attenuation is also encountered in the monochromatic simulations.  In the 15 mHz simulation, the 20$\%$ oscillation amplitude imposed at the upstream boundary is reduced to roughly 18$\%$ at (20,0,0) $R_E$ in the upstream solar wind.  In the 25 mHz simulation, the 40$\%$ oscillation amplitude imposed at the upstream boundary is reduced to roughly 16$\%$ at (20,0,0) $R_E$.  As the blue and green traces in \fig{filter} suggest, the input oscillation amplitude in the 10 mHz simulation is not significantly attenuated and remains at roughly 20$\%$ at (20,0,0) $R_E$.  Finally, we note that upstream $p_{dyn}$ driving with amplitudes in the 13-20$\%$ range is reasonable when compared with the observational studies discussed in \sec{intro} and is at the lower end of what has been reported.

The LFM simulation also does not contain a plasmaspheric model.  This simplification leads to plasma densities in the simulation inner magnetosphere ($r < 6.6$ $R_E$) that are lower than are typically observed in the real magnetosphere. For example, a nominal value for the number density near (5,0,0) $R_E$ in these LFM simulations is 1.0 particle/cm$^{3}$.  Using $B$ = 250 nT, the \alfven speed is then roughly 5,000 km/s in this region, which  is larger than in the real inner magnetosphere, where 1,000 km/s is a more typical value.  A related issue is the use of the ``Boris correction'' \citep{boris:70a} in the LFM simulation (see also \citet{lyon:04a}).  The Boris correction includes the perpendicular component of the displacement current in the $\bf{J} \times \bf{B}$ force of the  MHD  momentum equation, but with an artificially low speed of light, designated as $c_A$.  The \alfven wave phase speed is then given as: $v_{ph} = v_{A} (1+ v_{A}^{2}/c_{A}^{2})^{-\frac{1}{2}} \approx c_{A}$ when $v_{A} \gg c_{A}$.  Here, $v_A $ is the \alfven speed, $v_A = B/ \sqrt{\mu_0 \rho}$, where $B$ is the strength of the magnetic field, $\rho$ is the mass density and $\mu_{0}$ is the permeability of free space. The artificial speed of light, $c_A$,  used in these simulations is set to 5,500 km/s for computational efficiency.  The propagation speed of  \alfven waves is thus limited to $c_A$ in portions of the dayside magnetosphere.  In these LFM simulations, roughly 75$\%$ of the dayside magnetosphere is unaffected by the use of the Boris correction, in terms of closed flux volume.  It is only in regions of space where the spherical polar coordinate radius is less than $\approx$5 $R_E$ that the propagation speed of \alfven waves is limited to $c_A$.   We also note that if the plasma pressure plays a significant role, fast mode waves can exceed this speed and the code remains stable.  These issues will be important later when we compute the eigenfrequencies of individual field lines in the LFM simulation magnetosphere.

\section{Simulation Results \label{sec:simres}}

We begin by defining a quantity that will be used extensively, the {\it root-integrated power} ($RIP$) for a given time series:
\begin{equation*}
	RIP = \bigg [ \int_{f_{a}}^{f_{b}} P(f) df \bigg] ^{1/2}
\label{eq:rip}
\end{equation*}
Here, $P(f)$ is the power spectral density of a given time series and the integration is carried out over a given frequency band of interest, $[f_{a}, f_{b}]$.  Throughout this work,  we compute power spectral densities via the multi-taper method  \citep{thomson:82a,percival:93a}, which is essentially a windowed, averaged discrete Fourier transform spectral estimator \citep[e.g.][]{press:92a}.  For our spectral analyses, the only data manipulation preformed on the time series prior to the $PSD$ computation is a subtraction of the four-hour mean.  Note that root-integrated power is a measure of the oscillation amplitude in the frequency band, $[f_{a}, f_{b}]$.  For example, Parseval's theorem states that the $RMS$ amplitude of a given time series is equal to the $RIP$ when integrated over the entire frequency band $[0, f_{Ny}]$, where $f_{Ny}=1/2dt$ is the Nyquist frequency and $dt$ is the time sampling rate.  The Nyquist frequency in the 10 mHz and 15 mHz simulations is 16.67 mHz; in the 25 mHz and continuum simulations, it is 50 mHz.  Finally, note that if the units of the time series are nT, then the units of $P(f)$ are nT$^2$/Hz so that the units of $RIP$ are nT.

\fig{10sum}a shows the equatorial plane distribution of 10 mHz oscillations in the radial electric field driven by the monochromatic $p_{dyn}$  fluctuations (10 mHz simulation).  Throughout this work, we  analyze field components in cylindrical polar coordinates, ($r,\varphi,z$) \citep[e.g.][]{radoski:74a,hughes:94a}.  Root-integrated power ($RIP$) is shown on the color scale and is integrated over the {\it driving band}, which we define as a 1 mHz frequency band centered on the driving frequency.  Thus, for the 10 mHz simulation, the driving band is [9.5,10.5] mHz.   The panel shows a GSM equatorial plane cut of the driving band $RIP$ and the black contour is a snapshot of the magnetopause, determined by a field line tracing algorithm \citep{wiltberger:05a}.  The $x$ and $y$ axes are shown in black with 5 $R_E$ spaced tick marks and the sun is to the right.  The black disk at the origin  is the LFM inner boundary, located at  2.2 $R_E$.      Note that the bow shock is resolved  as the transition from  positive wave power in the magnetosheath to  zero wave power in the solar wind.  The bow shock intersects the $x$-axis at roughly 12 $R_E$ and the magnetopause intersects at roughly 9 $R_E$.  In \fig{10sum}a we see a strong amplitude oscillation in $E_r$ along $r$ $\approx$ 7 $R_E$ on the dayside.  Note that the $E_r$  oscillation amplitude along $r$ $\approx$ 7 $R_E$ is essentially zero at the noon meridian.  These features in $E_r$, shown here in the 10 mHz simulation, will be identified as the signature of a toroidal mode field line resonance.

Field line tracing algorithms for the LFM simulation \citep{wiltberger:05a} allow us to extract \alfven speed profiles along closed geomagnetic field lines in the simulations.  From these \alfven speed profiles, we can compute estimates to the natural oscillation frequency for a given field line. For example, the WKB estimate to the field line eigenfrequency is given by \citep{radoski:66a}:
\begin{equation}
	f_{n} = n \bigg [ 2 \int_{S}^{N} \frac{ds}{v_{A}(s)} \bigg] ^{-1} \quad \text{for } n=1,2,3,\cdots
\label{eq:efreq}
\end{equation}
Here, $n$ is the harmonic number of the FLR, $s$ is the arc length along the field line, and the integration is carried out from the southern field line foot point ($S$) to the northern foot point ($N$).  This estimate to the field line eigenfrequency is essentially the inverse of the travel time for an \alfven wave propagating along the magnetic field line to bounce off the reflecting ionospheres and return to its starting point \citep{radoski:66a,warner:79a}.  It is a good approximation for the higher harmonics (large $n$) and predicts a value about 20$\%$ higher than the actual fundamental mode eigenfrequency \citep{takahashi:84b}. The factor of 2 in the above equation is due to the aforementioned `bouncing' and due to the perfectly conducting ionospheric boundary condition assumed in FLR theory (e.g. \sec{intro}).   While the LFM ionospheric boundary condition described in \sec{method} results in finite conductivities in the model ionosphere,  the ionospheric conductivities are large enough that comparisons with perfect conductor FLR theory are appropriate \citep[e.g.][]{southwood:83a}.  A typical value for the dayside Pedersen conductivity in the LFM model ionosphere is 5 Siemens.

As noted above, the WKB approximation gives a somewhat crude  estimate to the fundamental field line eigenfrequency. Thus, we use an alternative method to estimate the local field line eigenfrequencies in the LFM simulation.  The toroidal mode \alfven wave eigenfrequencies are obtained using the toroidal mode \alfven wave dispersion relation from \citet{radoski:66b}  (their equation 25) with a perfect conductor boundary condition at the inner radial boundary of the LFM. Since there is some freedom of movement of the boundary (electric field mapped from the ionosphere), a more realistic boundary might be lower in altitude (leading to a longer field line length), so this boundary condition leads to an upper limit for the eigenfrequency.  The equation given by \citet{radoski:66b} is appropriate for a dipole magnetic field, but allows for an arbitrary dependence for the \alfven speed. For a given \alfven speed profile along an LFM field line (which is non-dipolar), we solve their equation using a dipole field line of the same length.  For closed LFM field lines on the dayside, the field line equatorial radius is typically ~5$-$10$\%$ smaller than the dipole field line of the same length.  This deviation from a dipole geometry is due to the compression of the LFM magnetosphere on the dayside.  Since the \alfven speed profile includes the magnitude of the magnetic field, the approximation that the field line is dipolar for the wave solution is only in error because of the scale factor associated with toroidal motion, $h_\varphi \propto r$. Examination of the wave equation of \citet{singer:81a}, which allows for arbitrary $h_\varphi$, shows that our approximation using the equation of \citet{radoski:66b} (with larger $h_\varphi$ than is realistic for the LFM field line) also leads to an upper limit for the eigenfrequency.  As we will see, this alternative estimate to the FLR eigenfrequency predicts eigenfrequencies closer to the simulated oscillation frequencies in the LFM magnetosphere when compared with the WKB estimate.  From here on, we will refer to this alternative estimate as the `dipole estimate.'

\begin{figure}
\begin{center}
 \includegraphics[scale=0.4]{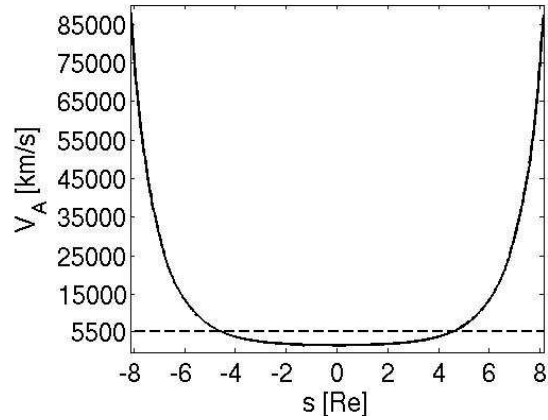}
\end{center}
 \caption{ \alfven speed profile along field line A in the 10 mHz monochromatic simulation (solid trace).  The dashed trace marks 5,500 km/s, the value of the Boris corrected speed of light in the LFM simulation.  Note that in the vicinity of the equatorial plane, the local \alfven speed is less than the speed of light in the LFM simulation. }
\label{fig:alfvenprof}
\end{figure}

The solid trace in \fig{alfvenprof} shows the \alfven speed profile along the field line that intersects the equatorial plane at ($x,y,z$) = (5.0, 5.0, 0), in the 10 mHz simulation.  In the $E_r$ panel in \fig{10sum}a, this field line intersects the center of the region of enhanced wave power in the afternoon sector.  In cylindrical coordinates, the location that this field line intersects the equatorial plane is  ($r,\varphi,z$) = (7.1, 45$^\circ$, 0).  In \fig{alfvenprof}, \alfven speed is plotted on the vertical scale from 0 to 90,000 km/s and distance along the field line, $s$, is plotted along the horizontal scale from $-8.1$ to 8.1 $R_E$.  The equatorial plane is located at $s$ = 0 and $s >$ 0 is defined to be the northern hemisphere.  Note that the \alfven speed is a minimum in the equatorial plane and increases towards the field line foot points, due to the increasing strength of the magnetic field.  It is important to remember that closed geomagnetic field lines in the LFM simulation domain have their foot points at the inner boundary of the simulation (a spherical shell of radius $\approx$ 2 $R_E$).  The dashed line in the figure marks the value of 5,500 km/s, which is the Boris corrected speed of light value in these LFM simulations.  As the figure shows, the local \alfven speed is less than the Boris corrected value  within the vicinity of the equatorial plane.  Moreover, the local \alfven speed  only exceeds the Boris corrected value near the ionosphere where the magnetic field strength is very large.  As noted above, a significant portion of the LFM dayside magnetosphere is not affected by the Boris correction and \alfven waves are able to develop and propagate naturally through these regions.  For the field line under consideration in \fig{alfvenprof}, we compute a fundamental eigenfrequency of $f_1$ = 11.7 mHz using the dipole estimate.  The WKB estimate (\eq{efreq})  gives $f_1$ = 15.3 mHz, significantly larger than both the dipole estimate and the actual frequency of oscillation in $E_r$ at this location, 10 mHz (e.g. \fig{10sum}a).  The reasonable agreement between the dipole estimate eigenfrequency and the oscillation frequency suggests that the oscillation in $E_r$ is the signature of the fundamental mode FLR.  As we will see below, the mode structure along the field line supports this claim.  Note that the dipole estimate is larger than the simulated frequency of oscillation, which is consistent with our reasoning above that this estimate provides an upper bound.

\fig{10sum} shows a summary of results from the 10 mHz monochromatic simulation. A similar format will be used in subsequent figures to show summaries of  the 15 mHz and 25 mHz simulation results.  As described above, \fig{10sum}a shows $E_r$ driving band $RIP$ in the equatorial plane. \fig{10sum}b shows $E_r$ driving band $RIP$ in the 15 LT (i.e. $\varphi$ = 45$^\circ$) meridional plane.  The black vertical axis is the $z$-axis and the horizontal axis lies in the equatorial plane and measures radial distance along the 15 LT meridian.  The tick marks are spaced at 5 $R_E$ on both axes.  In \fig{10sum}b, the white trace is the geomagnetic field line that intersects the equatorial plane at $r$ = 7.1  $R_E$ on the 15 LT meridian. This same field line  was used above to compute the local FLR eigenfrequency at ($r,\varphi,z$) = (7.1, 45$^\circ$, 0).  The black trace in the panel is the last closed field line in the 15 LT meridional plane.  \fig{10sum}b shows that the region of enhanced $E_r$ wave power in the afternoon LT sector of the equatorial plane (e.g. \fig{10sum}a) also extends out of the equatorial plane $\approx$ 2-3 $R_E$ in the positive and negative $z$-directions.  The region of enhanced $E_r$ wave power in the morning LT sector of the equatorial plane exhibits a similar structure out of the equatorial plane, though not shown here. The color scales  in \fig{10sum}a and \fig{10sum}b are the same and range from 0 to 5 mV/m.  In \fig{10sum}c, we plot $B_{\varphi}$ driving band $RIP$ in the 15 LT  meridional plane, i.e. the same plane as \fig{10sum}b.  $B_{\varphi}$ exhibits maxima at the low-altitude (ionospheric) ends of the field line and a node at the equator, as would be expected for a fundamental mode FLR. The last closed field line in the 15 LT meridional plane is again shown in black in \fig{10sum}b.  The same field line from \fig{10sum}b is shown in white and from here on we will refer to this field line as ``field line A'' in the 10 mHz simulation.  \tbl{fllocs} lists the location where field line A in the 10 mHz simulation intersects the equatorial plane, in both Cartesian and cylindrical polar coordinates.

\fig{10sum}d shows the $E_r$ power spectral density along the 15 LT meridian in the 10 mHz simulation.  Radial distance along 15 LT is plotted on the horizontal scale from 2.2 to 10.5 $R_E$ and frequency is plotted on the vertical scale from 0 to 16.5 mHz.  $E_r$ power spectral density ($PSD$) is plotted on the color scale from 0 to 80,000 (mV/m)$^2$/Hz.  The white shaded region near $r$ = 10 $R_E$ represents the approximate location of the magnetopause during the four hours of simulation time.   The solar wind dynamic pressure fluctuations result in magnetopause motion and this motion is represented by the extent of the shaded region. We note that the grid resolution here is sufficient to resolve this motion, with the motion occurring across roughly 2-3 grid cells.  The white dashed traces in the plot show the radial FLR eigenfrequency profiles derived from  the dipole estimate described above.  These four dashed traces correspond to  the fundamental mode ($n$ = 1) and the second through fourth harmonics ($n$ = 2, 3, and 4).  We also show the $n$ = 1 profile given by the WKB estimate (solid trace).  Note that the WKB estimate is roughly 20$\%$ larger than the dipole estimate, which is consistent with the aforementioned studies \citep[e.g.][]{takahashi:84b}.  Finally, in \fig{10sum}e, we show driving band $RIP$ profiles along field line A, for $E_r$ (solid blue trace) and $B_{\varphi}$ (solid green trace).  Distance, $s$, along field line A is shown on the horizontal scale and ranges from $-$8.1 to 8.1 $R_E$.   The equatorial plane is located at $s$ = 0 and $s >$ 0 is defined to be the northern hemisphere.  $E_r$ driving band $RIP$ is shown on the left vertical scale and ranges from 0 to 6.5 mV/m; $B_{\varphi}$ driving band $RIP$ is shown on the right vertical scale and ranges from 0 to 9 nT.  Note that these $RIP$ profiles are extracted along the white field line in \fig{10sum}b and \fig{10sum}c  and can be compared with those two panels for consistency.  The dashed traces in panel e) and the relevant features in panels a)-e) will be discussed in detail in \sec{discuss}.

\fig{15sum} and \fig{25sum} show a summary of results from the 15 mHz and 25 mHz monochromatic simulations, respectively.  A similar format to \fig{10sum} is used.  The main difference is that in the 15 mHz simulation, we consider the $RIP$ profiles along two field lines, as opposed to just  one in the 10 mHz simulation.  The locations where these two field lines intersect the equatorial plane are shown in \tbl{fllocs} and the two field lines are traced in white in \fig{15sum}b and \fig{15sum}c.  In the 25 mHz simulation, we again consider the $RIP$ profiles along two field lines.  The locations where these two field lines intersect the equatorial plane are shown in \tbl{fllocs} and the two field lines are traced in white in \fig{25sum}b and \fig{25sum}c.  For the 15 mHz and 25 mHz simulations, the $RIP$ plotted in the respective figures is integrated over the driving band.  The driving band in the 15 mHz simulation is 14.5-15.5 mHz and, in the 25 mHz simulation, is 24.5-25.5 mHz.

\fig{BBsum} shows the radial profile of $E_r$ $PSD$ plotted along the 15 LT meridian in the continuum simulation.  The dipole estimate to the FLR eigenfrequency profiles are overlaid in white dashed traces and the $n$ = 1 WKB estimate is again shown for comparison (solid white trace).  We note that there is no driving band in the continuum simulation as the upstream driving contains a continuum of ULF fluctuations in the 0-25 mHz band.  However,  a particular frequency can be chosen and qualitatively similar plots to the monochromatic simulation results can be produced.  For example, to consider fluctuations at 10 mHz in the continuum simulation, we could plot $RIP$ integrated over [9.5,10.5] mHz.  These plots reveal qualitatively similar features to the 10 mHz monochromatic simulation results in \fig{10sum}a-c (not shown here).  Similarly,  profiles of $E_r$ and $B_{\varphi}$ $RIP$ (integrated over [9.5,10.5] mHz) along a field line that intersects the equatorial plane in the continuum simulation at   ($r,\varphi,z$) = (7.1, 45$^\circ$, 0) look qualitatively similar to the profiles shown in \fig{10sum}e (not shown here).

\section{Discussion \label{sec:discuss}}

We now provide a detailed discussion of the simulation results presented in the previous section.  In \sec{evidence}, we interpret the simulation results as the signatures of toroidal mode field line resonances.  In \sec{polarization}, we show that the simulated  toroidal mode FLRs exhibit polarization signatures similar to those  seen in observations.  Finally, in \sec{energy}, we briefly consider the Poynting vector and discuss the relevant mechanisms for energy dissipation in the system.

\subsection{Evidence of Field Line Resonances \label{sec:evidence}}

We begin by considering the results from the 10 mHz monochromatic simulation, shown in \fig{10sum}.  We interpret the oscillations in $E_r$ and $B_{\varphi}$ on the dayside as signatures of toroidal mode field line resonances.  As discussed in \sec{intro}, the dominant oscillating field components for a toroidal mode FLR are expected to be $E_r$ and $B_{\varphi}$ in the approximate dipolar geometry of the dayside simulation region. We note here that the  `poloidal mode' and `toroidal mode' terminology  is strictly  valid only for an axisymmetric, dipole magnetosphere.  Portions of the real magnetosphere are certainly non-dipolar, particularly the nightside and the magnetotail.  Similarly, the LFM magnetosphere deviates from a dipolar configuration on the nightside \citep{lyon:04a}.  However, in the dayside LFM inner magnetosphere ($r <$ 10 $R_E$) in these simulations, the geometry is roughly dipolar.  We have confirmed this feature of the simulated magnetosphere by comparing contours of constant magnetic field  in the  equatorial plane  with circles and finding good agreement between the two on the dayside, within a tolerance of 0.5 $R_E$.  Thus, we use the terminology `toroidal mode' without hesitation and assume that oscillations in the $B_{\varphi}$ and $E_r$ LFM field components are an accurate representation of a toroidal mode oscillation on the dayside.

\begin{figure}[h]
\begin{center}
 \includegraphics[scale=0.22]{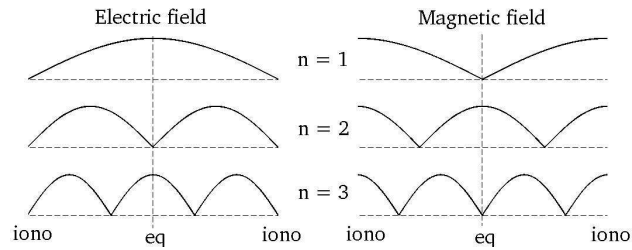}
\end{center}
 \caption{Oscillation amplitude schematics for resonant standing waves on straight magnetic field lines bounded by perfect conductors representing  geomagnetic field lines bounded by the ionosphere.  Electric and magnetic field oscillation  amplitudes are shown for the fundamental mode and the second and third harmonics.  To facilitate the comparisons with our simulation results, oscillation amplitude is plotted as a non-negative quantity.  The field line foot points are assumed to lie in a perfectly conducting ionosphere (labeled `iono').  The location of the geomagnetic equatorial plane is indicated by a vertical dashed line in each column (labeled `eq'). }
\label{fig:cartoon}
\end{figure}

Returning to the 10 mHz simulation results, the radial profile of $E_r$ wave power along the 15 LT meridian, shown in \fig{10sum}d, reveals an enhancement in 10 mHz wave power near $r$ = 7.1 $R_E$.  The radial profile of the fundamental mode field line eigenfrequency ($n$ = 1, dashed line) intersects this region of enhanced wave power.  This suggests that the 10 mHz oscillation in $E_r$ near $r$ = 7.1 $R_E$ on the 15 LT meridian is the signature of the fundamental toroidal mode FLR.  The profiles of $E_r$ and $B_{\varphi}$ $RIP$ along field line A (\fig{10sum}e), which intersects the 15 LT meridian at $r$ = 7.1 $R_E$, support this conclusion.  The standing wave schematics for the fundamental mode FLR, shown in \fig{cartoon}, indicate that in the equatorial plane,  the electric field has an antinode and the magnetic field has a node.  This structure of the field line oscillation amplitude profiles is evident in \fig{10sum}e.  Out of the equatorial plane along the field line, the  $B_{\varphi}$ oscillation amplitude (solid green) increases towards the field line foot points at the simulation inner boundary.  This profile is consistent with the $n$ = 1 schematic shown in \fig{cartoon}, where the magnetic field has antinodes at the field line foot points.  In addition,  the amplitude of the oscillation in $E_r$ (solid blue trace in \fig{10sum}e) decreases monotonically following the field line away from the equator until nodes are reached near $s$ = $\pm$4.5 $R_E$.  These two nodes are consistent with an $n$ = 1 FLR, though this is not obvious when comparing with \fig{cartoon}.

The additional nodes in the $E_r$ profiles are due to the coordinate system chosen for analysis, the curvature of the magnetic field lines and the effect of scale factors.  Field line curvature ensures that the  scale factors perpendicular to the magnetic field are not unity and change along the field line. In dipole coordinates, the perpendicular scale factor, $h_r$, multiplied by the perpendicular electric field, $E_r$, has a discrete harmonic solution along the field \citep{allan:79a,allan:79b,ozeke:04a}.  The choice of  cylindrical polar coordinates, ($r,\varphi,z$), also plays a role in the additional nodes in $E_r$.  In this coordinate system, the $r$-direction points outward from the cylindrical (i.e. dipole) axis.  Where the magnetic field line crosses the equatorial plane, the direction associated with the $r$-component is perpendicular to the magnetic field line (recall there is no dipole tilt in these simulations).  But as one moves out of the equatorial plane along field line A (e.g. \fig{10sum}b), a point is reached on the field line  where the (cylindrical) radial direction is tangent to the magnetic field direction.  Thus, in cylindrical polar coordinates, there must be a node in $E_r$ at this location (because ${\bf E} = - {\bf v} \times {\bf B}$ is always perpendicular to the equilibrium magnetic field, ${\bf B}$). The schematics shown in \fig{cartoon} do not show these two additional nodes in the  $n$ = 1 electric field profile because the schematics are drawn for a uniform plasma permeated by straight magnetic field lines. 

To verify that the FLR mode structure profiles reproduced by the LFM simulation are consistent with FLR theory, we use the same method that was used in \sec{simres} to compute the dipole estimate to the field line eigenfrequency.  Here, the theoretical mode structure profiles are obtained from the toroidal \alfven wave equations of \citet{radoski:66b} and the fields are projected in cylindrical polar coordinates.  Again, we use the \alfven speed profile from the LFM simulation along a dipole field line of the same length as the LFM field line.  The resulting mode structure profiles for the fundamental toroidal mode FLR are shown as dashed traces in \fig{10sum}e for $B_{\varphi}$ (green dashed) and $E_r$ (blue dashed).  The theoretical profiles obtained from the dipole dispersion relation  show very good agreement with the profiles reproduced by the LFM simulation.  We emphasize that the schematics shown in \fig{cartoon} are the ``standard'' picture of FLR oscillations often encountered in the literature \citep[e.g.][]{southwood:83a}.  However, we have shown here that the FLR mode structure is dependent upon the coordinate system chosen and care must be exercised when comparing simulations (and observations) with theory.  Field-aligned electric field mode structure profiles plotted in the $(\hat{\varphi} \times {\bf B} / \left\vert \hat{\varphi} \times {\bf B} \right\vert)$-direction, where {\bf B} is the equilibrium magnetic field, would not show these additional nodes. Note that the $\varphi$-variable in cylindrical polar coordinates is the same coordinate  in spherical polar and dipole coordinates.  Thus, the $B_{\varphi}$ profile shown in \fig{10sum}e will be consistent with the schematics shown in \fig{cartoon}, which are often envisioned in one of these two systems. The simulation results shown in \fig{10sum} provide convincing evidence to support the claim that the $n$ = 1 toroidal mode FLR is excited in the 10 mHz monochromatic simulation.

We now turn our attention to the results from the 15 mHz monochromatic simulation, shown in \fig{15sum}.  The radial profile of $E_r$ wave power along the 15 LT meridian (\fig{15sum}d) shows a strong enhancement in 15 mHz wave power near $r$ = 6.5 $R_E$ and a secondary,  weaker enhancement near $r$ = 9.2 $R_E$ (see also \fig{15sum}a and \fig{15sum}b).  In \fig{15sum}d, note that the $n$ = 1 field line eigenfrequency profile intersects the region of enhanced wave power near $r$ = 6.5 $R_E$ and that the $n$ = 3 profile intersects the enhanced wave power  near $r$ = 9.2 $R_E$.  This interpretation suggests that both the $n$ = 1 and $n$ = 3 toroidal mode FLRs are excited in the 15 mHz simulation.  The field line mode structure profiles shown in \fig{15sum}e and \fig{15sum}f support this conclusion.  The $B_{\varphi}$ mode structure along field line A (\fig{15sum}e), which intersects the equatorial plane at $r$ = 6.5 on the 15  LT meridian, shows a node in the equatorial plane and antinodes near the field line foot points.  This mode structure is consistent with the $n$ = 1 schematic for the magnetic field shown in \fig{cartoon}. In addition, the $E_r$ mode structure along field line A  shows an antinode in the equatorial plane,  nodes at the field line foot points, and two additional nodes near $s$ = $\pm$4 $R_E$.  As was the case for the 10 mHz simulation, this $E_r$ profile is consistent with an $n$ = 1 toroidal mode FLR, when considered in cylindrical polar coordinates and the effects due to scale factors are considered (e.g. \fig{10sum}e).  Thus, the available evidence indicates that the fundamental  toroidal mode FLR is excited in the 15 mHz simulation along field lines that intersect the equatorial plane near $r$ = 6.5 $R_E$ on the 15 LT meridian.

The $B_{\varphi}$ and $E_r$ mode structure profiles along field line B in the 15 mHz simulation (\fig{15sum}f) are consistent with the $n$ = 3 toroidal mode FLR interpretation.  The $B_{\varphi}$ profile again has a node in the equatorial plane but also has two additional nodes near $s$ = $\pm$8 $R_E$ along the field line.  This profile is consistent with the $n$ = 3 schematic for the magnetic field shown in \fig{cartoon}, where a node occurs in the equatorial plane, with an additional node between the equatorial plane and each field line foot point. In addition, the $E_r$ mode structure  along field line B   shows an antinode in the equatorial plane, nodes at the field line foot points, and additional nodes near  $s$ = $\pm$7 $R_E$ and  $s$ = $\pm$4 $R_E$. As was the case for the fundamental radial electric field oscillation, this mode structure is consistent with an  $n$ = 3 toroidal mode FLR, when considered in cylindrical polar coordinates and the effects due to scale factors are considered (not shown here).   Thus, the $n$ = 3 toroidal mode FLR is excited in the 15 mHz simulation along field lines that intersect the equatorial plane near $r$ = 9.2 $R_E$ on the 15 LT meridian.  Finally, we note that there are very faint features near 3.5 mHz in the $E_r$ spectrum along the 15 LT meridian (\fig{15sum}d).  These features are most likely due to aliasing \citep[e.g.][]{press:92a} and are very small amplitude (roughly an order of magnitude less than the 15 mHz oscillations).

We now consider the results from the 25 mHz monochromatic simulation, shown in \fig{25sum}.  The radial profile of $E_r$ wave power along the 15 LT meridian (\fig{25sum}d) shows enhancements in wave power near $r$ = 5.8 and 8.3 $R_E$.  The $n$ = 1 eigenfrequency profile intersects the region of enhanced wave power near   $r$ = 5.8 $R_E$, which suggests that this feature is the signature of the $n$ = 1 toroidal mode FLR.  The  $B_{\varphi}$ and $E_r$ mode structure  along field line A (\fig{25sum}e) supports this conclusion for the reasons discussed above.  Moreover,  the $n$ = 3 eigenfrequency profile intersects the region of enhanced wave power near   $r$ = 8.3 $R_E$, which suggests that this feature is the signature of the $n$ = 3 toroidal mode FLR.  Again, the $B_{\varphi}$ and $E_r$ mode structure  along field line B (\fig{25sum}f) supports this conclusion.  We also note the possible signature  of the $n$ = 5 FLR  in  \fig{25sum}b and \fig{25sum}c, in the region of space between field line B and the last closed field line.  Finally, we note that the FLR oscillation amplitudes in the three monochromatic simulations are the strongest in the 10 mHz simulation and the weakest in the 25 mHz simulation.  This variation is a result of the filtering/attenuation issue described in \sec{method}, where the attenuation of the oscillation amplitudes in the upstream dynamic pressure fluctuations becomes more pronounced at higher frequencies.

\begin{figure}
\begin{center}
 \includegraphics[scale=1.6]{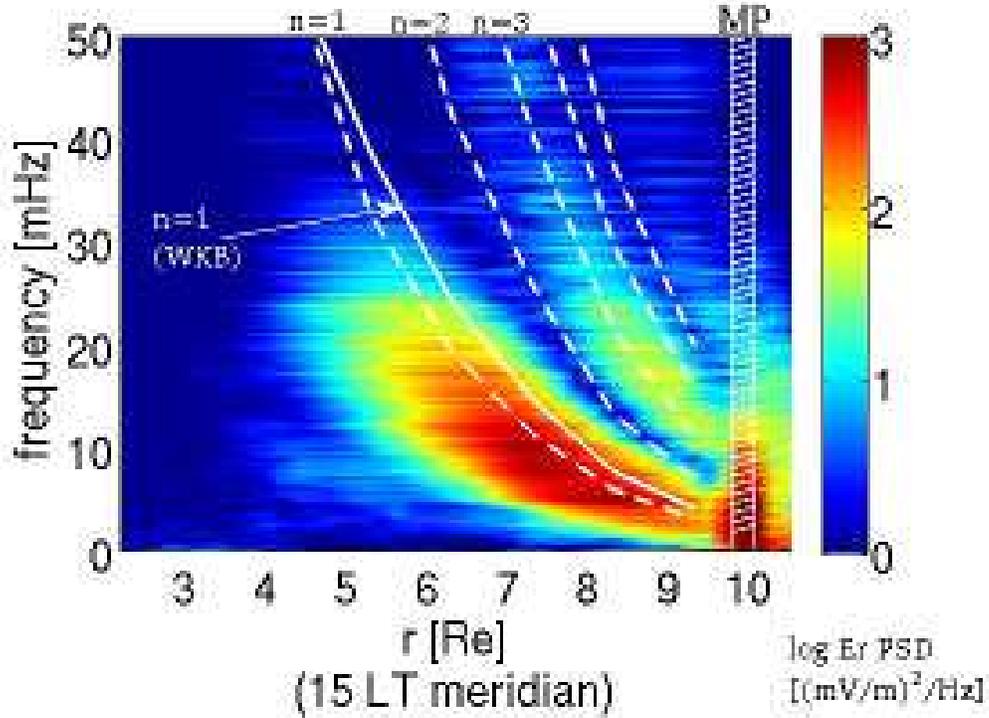}
\end{center}
 \caption{Summary of results from the continuum simulation showing the radial profile of $E_r$ power spectral density along the 15 LT meridian, on a logarithmic color scale.  The dashed  white traces  are the field line eigenfrequency profiles (dipole estimate) and the location of the magnetopause is indicated by the white shaded region near 10 $R_E$.  The $n$ = 1 WKB eigenfrequency estimate is also shown for comparison (solid white trace). Note the continuous spectrum of the $n$ = 1 and $n$ = 3  toroidal mode field line resonances excited along the 15 LT meridian. }
\label{fig:BBsum} 
\end{figure}

We now take up the results from the continuum simulation, shown in \fig{BBsum}.  The radial profile of  $E_r$ wave power along the 15 LT meridian shows remarkable agreement with the FLR eigenfrequency profiles (dashed traces).  The $n$ = 1 and $n$ = 3 eigenfrequency traces coincide with the regions of enhanced wave power along the 15 LT meridian.  This coincidence suggests that the continuous spectra of $n$ = 1 and $n$ = 3 FLRs are excited in the continuum simulation.  Though only shown here at the 15 LT meridian, the $n$ = 1 and $n$ = 3 FLRs are excited across all of the dayside, excluding near the noon meridian, as was the case in the monochromatic simulations.  Note that the $n$ = 2 and  $n$ = 4 eigenfrequency traces lie in between the  regions of enhanced wave power, suggesting that only odd mode number toroidal mode FLRs are excited in the continuum simulation.  This effect is easily explained by  considering the nature of the upstream forcing.  As discussed in \sec{method}, in the LFM simulation the solar wind  propagates as planar fronts towards the magnetosphere.  Furthermore, in these LFM simulations, the solar wind velocity has only an $x$-component, so that the planar fronts are parallel to the $yz$-plane and propagate in the negative $x$-direction.  As these idealized simulations are conducted with no dipole tilt, the planar fronts impact the subsolar, equatorial plane of the magnetosphere first, which lies on the positive $x$-axis. This impact leads to a situation where the forcing is symmetric about the equatorial plane, along closed field lines. Consequently, the perturbations in $E_r$ are symmetric and those in $B_{\varphi}$ are antisymmetric, with respect to the equatorial plane \citep{allan:85a,lee:89a} and only odd mode number toroidal mode FLRs can be excited.  We note that a similar situation was reported in the numerical simulations of  \citet{lee:89a,lee:91a} who modeled FLRs in a dipole geometry with impulsive  upstream forcing, which was also symmetric about the equatorial plane. Additionally, we note that the absence of the $n$ = 2 and $n$ = 4 harmonics can be understood by the fact that the overlap integrals to these modes are identically zero \citep[e.g.][]{southwood:86a,wright:92a}.  If a slight tilt were introduced to the dipole axis, there would be a weak coupling to the $n$ = 2 and $n$ = 4 harmonics.  The node in $E_r$ near the noon meridian is also due to the symmetry in the upstream forcing and a node in $\partial B_z / \partial \varphi$, which is proportional to the linear magnetic pressure gradient in the azimuthal direction \citep{rickard:94a,rickard:95a}.  In the MHD momentum equation, this quantity is responsible for forcing the $\varphi$-component,  hence the node in $v_{\varphi}$ and $E_r$.  Similarly, the monochromatic simulation results (\fig{10sum}, \fig{15sum}, and \fig{25sum}) do not show any FLRs near noon or any even mode number FLR signatures.  The monochromatic simulations were also conducted with no dipole tilt and driven with  planar fronts propagating in the negative $x$-direction. It is interesting to note that observational studies of FLRs often report that the mode structure is statistically odd mode number \citep[e.g.][]{anderson:90a}.  Finally, note that the radial profile of $E_r$ $PSD$ along the 15 LT meridian in the continuum simulation (\fig{BBsum}) shows that multiple FLR harmonics can be excited on the same field line, which has been reported in observations \citep[e.g.][]{singer:79a}.

\subsection{Polarization Characteristics  \label{sec:polarization}}

\begin{figure}[h]
\begin{center}
 \includegraphics[scale=0.375]{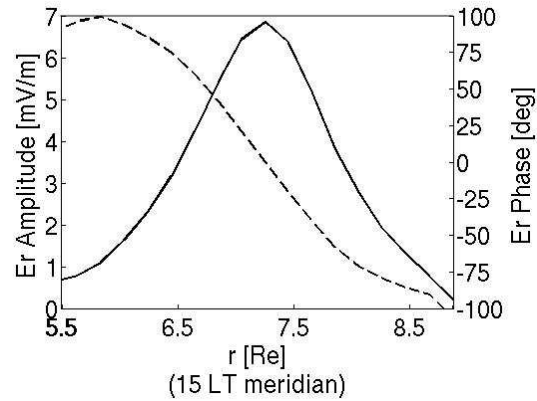}
\end{center}
 \caption{ Radial electric field amplitude (solid) and phase (dashed) for the fundamental mode FLR in the 10 mHz simulation.  Radial distance, $r$, along the 15 LT meridian is plotted on the horizontal scale from 5.5 to 8.9 $R_E$.  $E_r$ driving band $RIP$ (labeled `$E_r$ amplitude') is plotted on the left vertical scale from 0 to 7 mV/m and relative phase is plotted on the right vertical scale from $-$100 to 100 degrees.  Note the approximately 180$^{\circ}$ change in phase across the peak in wave power.}
\label{fig:powphase}
\end{figure}

Observations of ULF pulsations attributed to field line resonances often show an approximately 180$^{\circ}$ phase change radially across a peak in wave power \citep{samson:71a,walker:79a,ruohoniemi:91a}.  This observational signature is predicted by the FLR theory \citep{chen:74a,southwood:74a}.  The FLRs in the LFM simulation also exhibit this signature, as illustrated in \fig{powphase} for the fundamental mode FLR in the afternoon sector in the 10 mHz simulation (e.g. \fig{10sum}a near $r$ = 7.1 $R_E$).   The solid trace in \fig{powphase} shows the radial profile of the $E_r$ driving band $RIP$ along the 15 LT meridian. Radial distance, $r$, along the 15 LT meridian is shown on the horizontal scale from 5.5 to 8.9 $R_E$.  The $E_r$ driving band $RIP$ is shown on the left vertical scale (labeled `$E_r$ amplitude' in the figure) and ranges from 0 to 7 mV/m.  The solid trace in \fig{powphase} can be compared with \fig{10sum}a for consistency.  The dashed trace in \fig{powphase} shows the relative phase computed from the $E_r$ time series along 15 LT meridian via a fast Fourier transform method.  The $E_r$ time series at the peak in wave power serves as the reference time series (i.e. relative phase = 0) and relative phase is plotted on the right vertical scale from $-$100 to 100 degrees.  From the figure, we see  an approximately 180$^{\circ}$ change in phase radially across the peak in power, i.e. radially across the location of the FLR in the equatorial plane.  This signature of the simulated FLR, including the decrease in phase as $r$ increases, is consistent with both the theory \citep[e.g.][]{chen:74a} and the observations \citep[e.g.][]{ruohoniemi:91a} described above.  The $n$ = 1 and $n$ = 3 FLRs identified in the 15 mHz and 25 mHz simulations also show  phase changes radially across the respective peaks in wave power. We note, however, that the phase change is not always $\approx$ 180$^{\circ}$, as it is for the $n$ = 1 FLR in the 10 mHz simulation.  For example, for the $n$ = 1 and $n$ = 3 FLRs in the 15 mHz simulation, the phase changes are in the 160-200$^{\circ}$ range.  In the 25 mHz simulation, the phase changes are on the order of 100$^{\circ}$, which is significantly less than the phase change in the 10 mHz  simulation.  The deviations from the theoretical expectations may be due to the interaction between multiple resonances \citep{mcdiarmid:99a}, as there is only the fundamental  resonance in the 10 mHz simulation.  Also, as noted above, the continuum simulation does not show a radial peak in wave power along the 15 LT meridian unless power at a particular frequency (from the continuum of FLR frequencies) is considered  (e.g. \fig{BBsum}).  If power at a particular frequency is chosen for analysis, then we find a similar phase reversal across the peak in wave power (not shown here).

\begin{figure}
\begin{center}
 \includegraphics[scale=0.45]{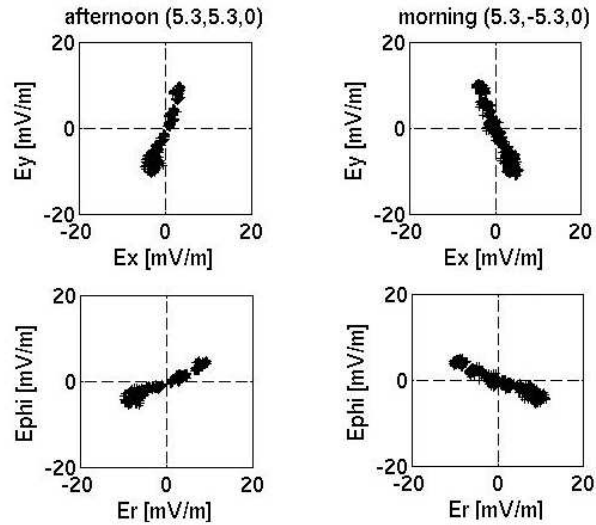}
\end{center}
 \caption{Electric field hodograms near the location of the fundamental mode FLR in the 10 mHz simulation.  The left column shows the hodograms in the afternoon sector and the right column shows the hodograms in the morning sector.  The top row shows the Cartesian coordinate electric field components, $E_x$ and $E_y$, and the bottom row shows the polar coordinate components,  $E_r$ and $E_{\varphi}$.  Note the linearly polarized oscillations at the locations of the resonance  and the polarization reversal across the noon meridian.}
\label{fig:polrev}
\end{figure}

Another feature of FLR observations is a change in the sense of polarization across local noon \citep{samson:71a,fukunishi:75a}.  \fig{polrev} shows hodograms of the electric field near the location of the fundamental mode FLR in the 10 mHz simulation (e.g. \fig{10sum}a near $r$ = 7.1 $R_E$).  The left column in the figure shows the electric field hodograms in the equatorial plane in the afternoon sector at {\bf x} = (5.3,5.3,0); the right column shows the hodograms in the morning sector at {\bf x} = (5.3,-5.3,0).  The top row shows the hodograms for the Cartesian coordinate electric field components, $E_x$ and $E_y$, where a `+' is plotted at each time step for the entire four hours of simulation time.  The bottom row shows the hodograms for the cylindrical polar coordinate electric field components, $E_r$ and $E_{\varphi}$. The only data manipulation done on the time series is a removal of the four-hour mean; the time series are not filtered.   Comparing the two hodograms in the top row, we see that the axis tilt in the electric field polarization has a positive slope in the afternoon sector and a negative slope in the morning sector.  This feature indicates a polarization reversal in the electric field oscillations across the noon meridian, a result consistent with FLR observations.  The FLR theory of \citet{chen:74a} and \citet{southwood:74a}  predicts that the polarization is linear near the peak in wave power, which is clearly evident in \fig{polrev}.  Moreover, the theory also predicts that the  polarization is elliptical for $L$-shells adjacent to the resonance, which is also reproduced in the LFM simulation (not shown here). Hodograms of the electric field components near the morning and afternoon sector locations of the FLRs in the 15 mHz and 25 mHz simulations show similar polarization reversals across local noon.

Several authors have noted that the observed polarization reversal across 12 LT can be explained by assuming that the compressional energy that drives the FLRs originates in Kelvin-Helmholtz waves near the magnetopause \citep{samson:71a,walker:79a,southwood:83a}.  However, any disturbance whose origin lies in the solar wind and propagates around the magnetosphere with the solar wind can also produce the observed change in polarization across local noon \citep{hughes:94a}.  Thus, the observed polarization change  does not necessarily implicate the Kelvin-Helmholtz instability for FLR excitation; it merely implicates the solar wind.  Other authors \citep{yumoto:83a,rostoker:87a} have argued against a Kelvin-Helmholtz mechanism for FLR excitation on a variety of grounds.  The field line resonances in this study are ultimately driven by solar wind dynamic pressure fluctuations, which results in the polarization reversal across local noon.

\subsection{Energy Considerations \label{sec:energy}}

We now consider the flux of energy near the FLR in the LFM simulation. The energy flux is examined by considering the Poynting vector, {\bf S}:
\begin{equation*}
	{\bf S} = \frac{ {\bf e} \times {\bf b}  }{\mu_0}
\label{eq:poynt}
\end{equation*}
where {\bf e} is the perturbed electric field and {\bf b} is the perturbed magnetic field.  The perturbed magnetic field is defined as ${\bf b} = {\bf B} - \text{mean}({\bf B})$, where {\bf B} is the magnetic field and the mean is calculated over the four hours of simulation time (similarly for the perturbed electric field).  In fact, a more useful quantity is the time averaged Poynting vector, $\langle${\bf S}$\rangle$:
\begin{equation*}
	\langle {\bf S}({\bf x}) \rangle = \frac{1}{T}\int_{0}^{T} {\bf S}({\bf x},t) dt
\label{eq:poynt2}
\end{equation*}
where $T$ is the period of the oscillation.  The time averaged Poynting vector gives the net energy flux over one period of oscillation, as opposed to the instantaneous direction of energy flux  given by {\bf S}.  As we are considering the Poynting vector, our discussion of energy flow only accounts for electromagnetic power, i.e. we do not consider the contributions from the flux of kinetic energy or thermal energy \citep[e.g.][]{kouznetsov:95a}.

\begin{figure}
\begin{center}
 \includegraphics[scale=0.4]{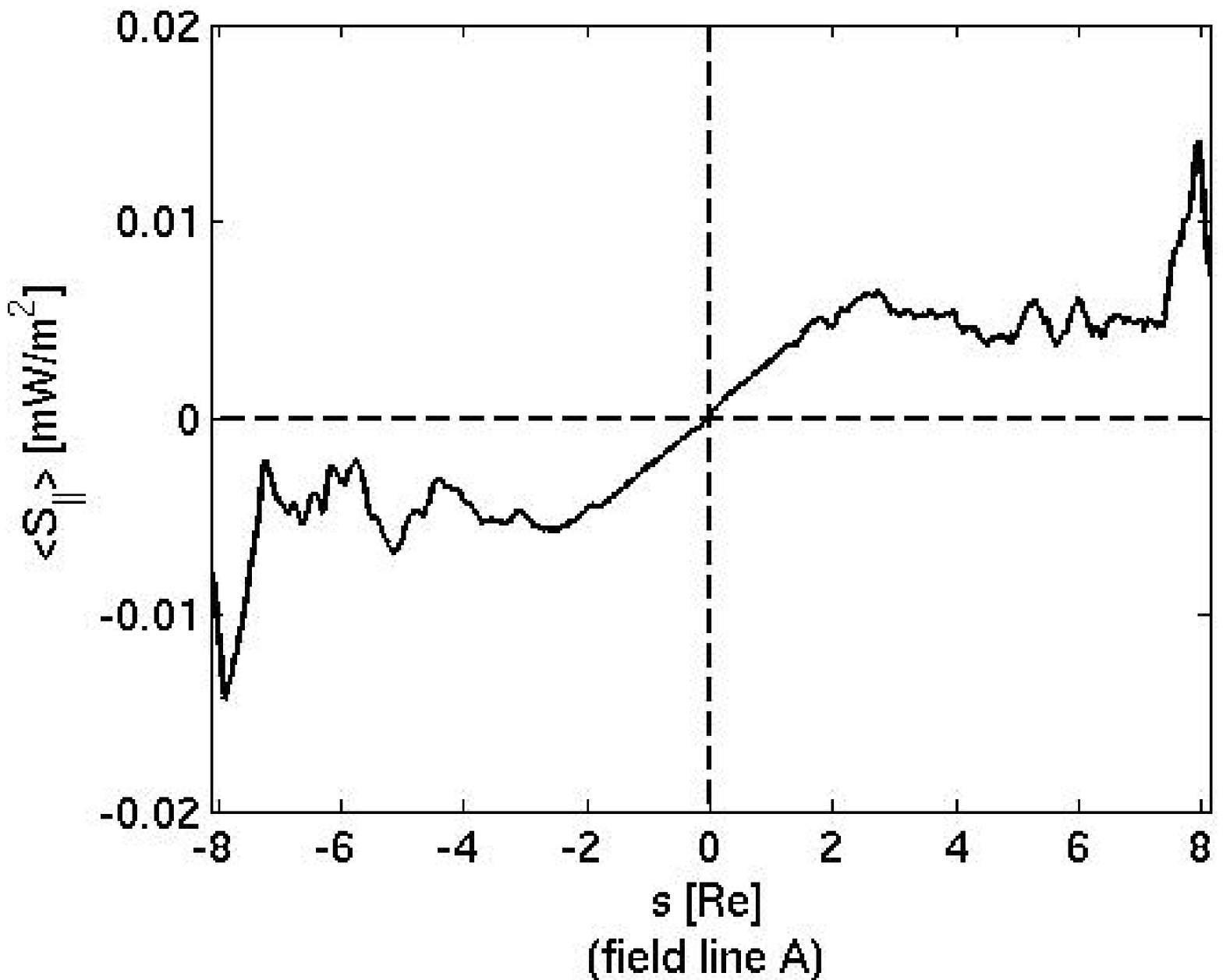}
\end{center}
 \caption{Parallel component of the time averaged Poynting vector  along field line A in the 10 mHz simulation.  Note the net  flux of energy  out of the equatorial plane, towards the northern and southern ionospheres.  }
\label{fig:poyntFL}
\end{figure}

\fig{poyntFL} shows the parallel component of the time averaged Poynting vector, $\langle S_{\parallel} \rangle$, along field line A in the 10 mHz simulation.  We define a positive (negative) value of $S_{\parallel}$ as the component of {\bf S} parallel (antiparallel) to {\bf B}.  Distance, $s$, along field line A is plotted on the horizontal scale from $-$8.1 to 8.1 $R_E$  and $s >$ 0 is defined to be the northern hemisphere.    The profile of  $\langle S_{\parallel} \rangle$ shows that, near the equatorial plane, a net flow of energy is directed from the equatorial plane towards the northern and southern ionospheres.  The direction is consistent with the development of a resonant field line oscillation forced  at the equatorial plane.

In the three monochromatic simulations, the monochromatic fluctuation in the upstream dynamic pressure drives compressional waves on the dayside magnetosphere (not shown here).  These compressional waves energize toroidal mode field line resonances at locations where the frequency of the compressional source matches a local field line eigenfrequency.  The open magnetospheric configuration in the LFM simulation allows for compressional energy to be lost into the nightside magnetosphere.  This geometric configuration is a significant advance over the closed magnetospheric numerical  simulations of the past \citep[e.g.][]{lee:89a,lee:91a},  is a more realistic representation of the solar wind-magnetosphere-ionosphere system than previous work in open geometries \citep{lee:99a,lee:04a}, and provides a new tool for studying resonant oscillations in the magnetosphere.

The controlled nature of the LFM numerical experiments described in this work allows us to probe deeper into the flow of energy through the coupled solar wind-magnetosphere-ionosphere system.  To this end, we ran the four simulations in this study with the upstream dynamic pressure fluctuations turned off after the four hour driving phase. The resonant field line oscillations die out quickly, within roughly 2-3 wave periods.  In the real magnetosphere, ionospheric dissipation may be the primary mechanism for FLR decay \citep{newton:78a}.  However, this dissipation does not appear to be the limiting mechanism in the version of LFM simulation used in this study; the simulation results suggest that the rapid decay of the FLRs is due to numerical dissipation.  To confirm this assertion, we have conducted several other LFM simulations, nearly identical to the 10 mHz simulation presented here.  In the first set of new simulations, we changed the solar 10.7 cm flux, which serves as an input to the ionospheric model (see \sec{method}), to 200$ \times \text{10}^\text{-22}$ J/m$^2$ and 300$ \times \text{10}^\text{-22}$ J/m$^2$ (it was 100$ \times \text{10}^\text{-22}$ J/m$^2$ in the 10 mHz monochromatic simulation).  The increase in solar 10.7 cm flux has the effect of increasing the Pedersen conductivity in the dayside ionosphere from its nominal value of roughly 5 Siemens in the 10 mHz simulation, to a value of roughly 8.5 and 10 Siemens in the two new simulations (near the foot points of the resonant field lines).  The full-width at half maximum (FWHM) of the $E_r$ amplitude in the equatorial plane, which is governed by the decay rate of the FLR, does not vary significantly across the three simulations and has a value of roughly 1.6 $R_E$.  As a second set of tests, we ran the LFM simulation coupled to a simpler ionospheric model, one in which the Pedersen conductivity is maintained at a constant value \citep{fedder:95a}.  We ran three simulations in which the upstream driving conditions were identical to the 10 mHz simulation but where the constant value of the ionospheric Pedersen conductivity was maintained at 5, 10, and 20 Siemens, respectively.  Again, the FWHM of the $E_r$ amplitude in the equatorial plane did not vary significantly across the three simulations and had a value of roughly 1.7 $R_E$.  In either set of tests, the FWHM of the resonant oscillations did not change significantly when the ionospheric Pedersen conductivity was changed.  Thus, in the LFM simulation, the FLR decay rates, which are related to the FWHM, are not significantly affected by the value of the Pedersen conductivity in the ionosphere.  This result implies that the finite ionospheric conductance is not the dominant sink of FLR power in the LFM and we  suggest that the rapid decay of the FLRs is primarily through numerical dissipation.

Finally, we note that there are roughly six grid cells across the FWHM in these LFM simulations.  If the FLRs are not fully resolved in the radial direction, there can be an unphysical leakage of wave energy from the FLR into the fast compressional mode \citep{stellmacher:97a}. Thus, the rapid decay of the FLRs in these simulations may not be entirely due to numerical dissipation.  In addition, there could be damping due to the fact that the LFM grid is not magnetically aligned.  In the non-Cartesian, distorted spherical mesh used in the LFM simulation \citep{lyon:04a}, resonant field lines thread across grid cells, into neighboring grid cells that have field lines magnetically connected to different $L$-shells.  Thus, wave energy from the resonant $L$-shell can leak into neighboring, non-resonant $L$-shells.   This would also be the case for other global MHD models, such as BATS-R-US \citep{gombosi:03a} and OpenGGCM \citep{raeder:01a}, which both use a Cartesian grid.

We acknowledge that the decay rates/FWHMs are probably not a realistic representation of FLRs in the real magnetosphere.  However, the main results of this study are: that FLRs can be driven by dynamic pressure fluctuations and that the LFM simulation can reproduce FLRs in a realistic solar wind-magnetosphere-ionosphere geometry with global FLR characteristics consistent with observations.  We emphasize that none of these results depend critically on the decay rate of the FLRs, as we are continually driving the magnetosphere during the time periods in which the above results are obtained. The dissipation of energy in the LFM magnetospheric system warrants a more detailed investigation and this will be the subject of future work.  A related issue is the presence of cavity/waveguide mode oscillations in the LFM simulation and the nature of the coupling between the FLR and the cavity/waveguide mode.  The simulation results presented here suggest that the LFM is able to reproduce cavity/waveguide mode oscillations under these driving conditions (see also \citet{claudepierre:09a}).  Moreover, the open configuration of the LFM magnetosphere allows for the fast mode compressional energy to be lost downtail and the cavity behaves more like a waveguide \citep[e.g.][]{samson:92a,harrold:92a,wright:94a,rickard:94a}.  These issues related to the coupling of cavity mode/waveguide energy into the FLRs in the LFM simulation will also be the subject of future work.

\section{Summary and Conclusions \label{sec:conclu}}

In this work, the Lyon-Fedder-Mobarry (LFM) global, three-dimensional MHD simulation was used to investigate the role that solar wind dynamic pressure fluctuations play in the generation of magnetospheric ULF pulsations.   The LFM simulations were driven with idealized solar wind input parameters, where ULF fluctuations were introduced into the time series of solar wind dynamic pressure.  In three of the simulations, a monochromatic sinusoidal variation in the solar wind dynamic pressure was used to drive the magnetosphere.  In these simulations, the monochromatic fluctuation in the upstream dynamic pressure drove compression of the dayside magnetosphere. Resulting compressional waves energized toroidal field line resonances at locations where the frequency of the compressional source, ultimately solar wind pressure oscillations, matched a local field line eigenfrequency.  In the fourth simulation, a continuum of ULF frequencies was introduced into the upstream dynamic pressure fluctuations.  This configuration served as a quasi-broadband stimulus to the magnetosphere and resulted in the excitation of a continuous spectrum of toroidal mode field line resonances on the dayside.   The simulated field line resonances also showed several characteristics consistent with observations, including a phase reversal radially across the location of the resonance, a change in the sense of polarization across the noon meridian, and a net flux of energy out of the equatorial plane into the ionosphere.

The idealized solar wind input parameters used in this study were chosen to represent continuous driving of the magnetosphere.  In contrast with the driving in the continuum simulation, a discrete feature in the solar wind dynamic pressure could also have been used to drive the simulation.  A step function increase in the solar wind dynamic pressure would provide a true  broadband stimulus to the magnetosphere.  We have analyzed results from idealized LFM simulations where a large step function increase in the solar wind dynamic pressure  ($\Delta p / p \sim \text{5}$) was used to drive the magnetosphere.   The initial results suggest that the broadband stimulus can drive FLRs on the dayside, though the oscillations die out quickly, within 2-3 wave periods.  

\begin{figure}
\begin{center}
 \includegraphics[scale=1.8]{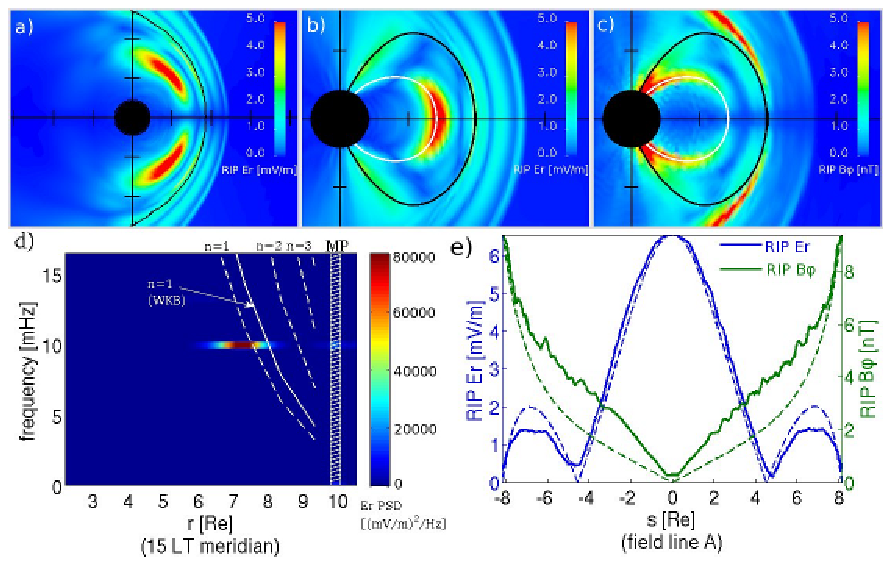}
\end{center}
 \caption{ Summary of results from the 10 mHz monochromatic simulation.  {\bf a)}  Equatorial plane distribution of driving band ([9.5,10.5] mHz) $E_r$ root-integrated power ($RIP$).   {\bf b)} Distribution of driving band $E_r$ $RIP$ in the 15 LT meridional plane.  The white field line is field line A in the 10 mHz simulation and the black field line is the last closed field line in the 15 LT meridional plane. Field line A intersects the equatorial plane at $r$ = 7.1 $R_E$ on the 15 LT meridian.  {\bf c)} Distribution of driving band $B_{\varphi}$ $RIP$ in the 15 LT meridional plane. The same two field lines from panel b) are shown. {\bf d)} Radial profile of $E_r$ power spectral density along the 15 LT meridian.  The dashed  white traces  are the field line eigenfrequency profiles (dipole estimate) and the location of the magnetopause is indicated by the white shaded region near 10 $R_E$.  The $n$ = 1 WKB eigenfrequency estimate is also shown for comparison (solid white trace).  {\bf e)} $E_r$ (solid blue) and $B_{\varphi}$ (solid green) mode structure profiles along field line A.  The two solid traces can be compared with panels b) and c), where field line A is shown in white.   The dashed traces are the theoretical mode structure profiles for fundamental toroidal mode FLR oscillations in $E_r$ and $B_{\varphi}$ and are discussed in greater detail in \sec{evidence}. Note the fundamental toroidal mode field line resonance excited near $r$ = 7.1 $R_E$ across the entire dayside, excluding near the noon meridian. }
\label{fig:10sum}
\end{figure}

\begin{figure}
\begin{center}
 \includegraphics[scale=1.8]{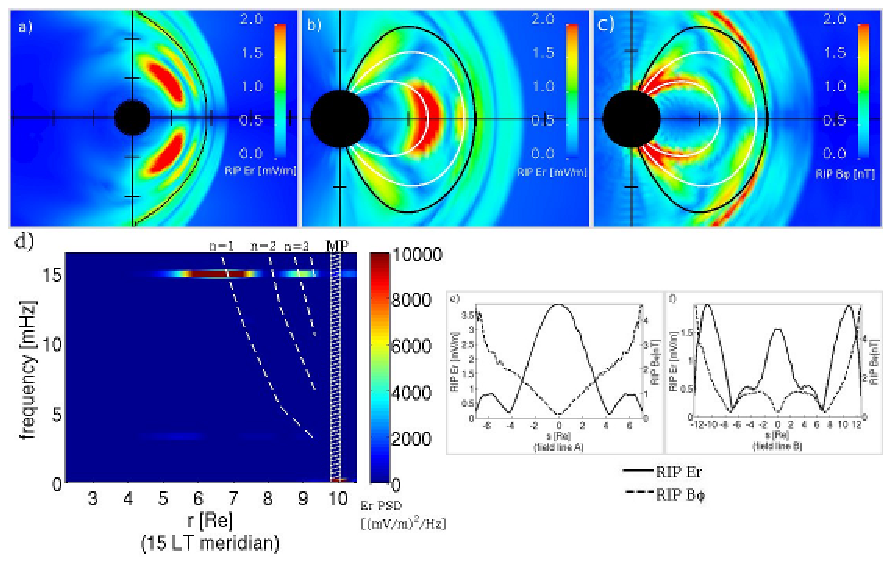}
\end{center}
 \caption{ Summary of results from the 15 mHz monochromatic simulation.  {\bf a)}  Equatorial plane distribution of driving band ([14.5,15.5] mHz) $E_r$ root-integrated power ($RIP$).  {\bf b)} Distribution of driving band $E_r$ $RIP$ in the 15 LT meridional plane.  The two white field lines are field lines A and B in the 15 mHz simulation and the black field line is the last closed field line in the 15 LT meridional plane. Field line A intersects the equatorial plane at $r$ = 6.5 $R_E$ on the 15 LT meridian and field line B intersects  at $r$ = 9.2 $R_E$.  {\bf c)} Distribution of driving band $B_{\varphi}$ $RIP$ in the 15 LT meridional plane. The same three field lines from panel b) are shown. {\bf d)} Radial profile of $E_r$ power spectral density along the 15 LT meridian.  The dashed  white traces  are the field line eigenfrequency profiles (dipole estimate) and the location of the magnetopause is indicated by the white shaded region near 10 $R_E$.  {\bf e)} $E_r$ and $B_{\varphi}$ mode structure profiles along field line A.  The two traces can be compared with panels b) and c), where field line A is shown in white.   {\bf f)} $E_r$ and $B_{\varphi}$ mode structure profiles along field line B.  The two traces can be compared with panels b) and c), where field line B is shown in white.    Note the $n$ = 1 and $n$ = 3  toroidal mode field line resonances excited across most of the dayside, near $r$ = 6.5 and 9.2 $R_E$, respectively. }
\label{fig:15sum}
\end{figure}

\begin{figure}
\begin{center}
 \includegraphics[scale=1.8]{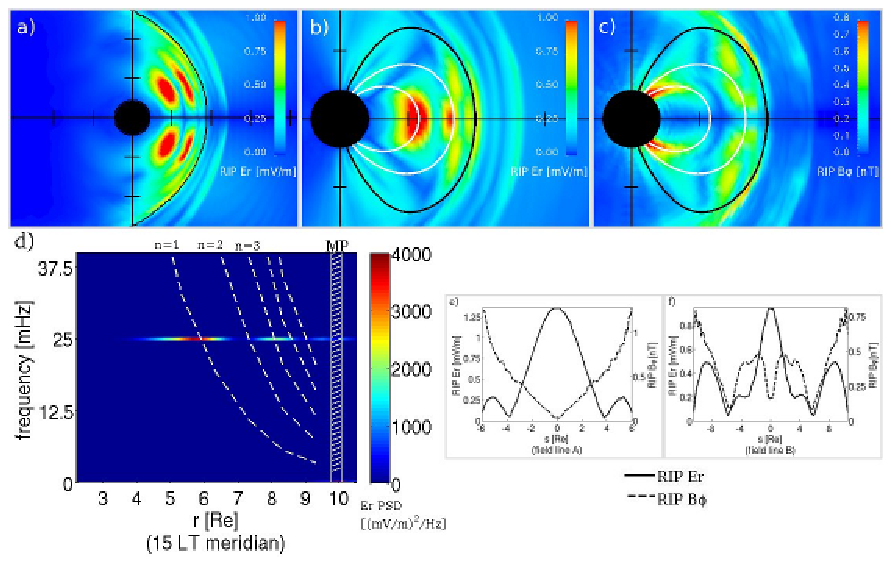}
\end{center}
 \caption{ Summary of results from the 25 mHz monochromatic simulation.  {\bf a)}  Equatorial plane distribution of driving band ([24.5,25.5] mHz) $E_r$ root-integrated power ($RIP$).  {\bf b)} Distribution of driving band $E_r$ $RIP$ in the 15 LT meridional plane.  The two white field lines are field lines A and B in the 25 mHz simulation and the black field line is the last closed field line in the 15 LT meridional plane. Field line A intersects the equatorial plane at $r$ = 5.8 $R_E$ on the 15 LT meridian and field line B intersects at $r$ = 8.3 $R_E$.  {\bf c)} Distribution of driving band $B_{\varphi}$ $RIP$ in the 15 LT meridional plane. The same three field lines from panel b) are shown. {\bf d)} Radial profile of $E_r$ power spectral density along the 15 LT meridian.  The dashed  white traces  are the field line eigenfrequency profiles (dipole estimate) and the location of the magnetopause is indicated by the white shaded region near 10 $R_E$.  {\bf e)} $E_r$ and $B_{\varphi}$ mode structure profiles along field line A.  The two traces can be compared with panels b) and c), where field line A is shown in white.   {\bf f)} $E_r$ and $B_{\varphi}$ mode structure profiles along field line B.  The two traces can be compared with panels b) and c), where field line B is shown in white.    Note the $n$ = 1 and $n$ = 3  toroidal mode field line resonances excited across most of the dayside, near $r$ = 5.8 and 8.3 $R_E$, respectively. }
\label{fig:25sum}
\end{figure}

The majority of the analyses in this study were done on the LFM fields near the 15 LT meridian in the afternoon sector.  However, the simulated FLRs are not localized near 15 LT and extend across the entire dayside, excluding the  12 LT region (e.g. \fig{10sum}a, \fig{15sum}a and \fig{25sum}a).  We also emphasize that the values of the resonant frequencies reported in this study should not be compared directly with observations.  The primary factor controlling the field line eigenfrequencies in the simulations is the \alfven speed profile along the field line (the length of the field line and the ionospheric boundary conditions are the other controlling factors).   The version of the LFM simulation code used in this study does not contain a plasmaspheric model, which leads to plasma densities in the inner magnetosphere  lower than those typically observed in the real magnetosphere.  The low values for the plasma density result in large values of the \alfven speed and large FLR eigenfrequency values on the LFM  dayside magnetosphere.   A new version of the LFM simulation, with a realistic plasmaspheric model, is currently under development and will be used in future studies of resonant ULF waves in the LFM.

The study of ULF pulsations presented in this work offers a new perspective of resonant ULF waves, when compared with previous numerical work.  The most obvious difference between this and previous work is the realistic magnetospheric geometry in the LFM simulation.  The numerical studies of \citet{lee:89a,lee:91a} and \citet{allan:86a} used a dipolar magnetospheric model and a hemi-cylindrical magnetospheric model, respectively.  Both of these model magnetospheres were closed and, thus, did not allow for a downtail loss of compressional wave energy into the nightside, which the LFM is able to reproduce in a realistic fashion. To the best of our knowledge, the only other study that has used a random/continuum driver to excite FLRs is the work of \citet{wright:95b}.  The differences between the \citet{wright:95b} study and the present study are again likely due to the different geometries used. Moreover, the realistic interaction of the solar wind with the magnetosphere in the LFM simulation allows for the natural development of the magnetopause, bow shock, and magnetosheath.    In addition, in the numerical work of \citet{lee:89a,lee:91a} and \citet{allan:86a}, a particular value for the azimuthal mode number, $m$, must be specified in their models.  In the LFM simulations presented in this work, the azimuthal mode structure of the resonant oscillations falls out naturally  from the model.  Spectral analyses of the equatorial plane electric field components shows that  $m \leq 6$ for the oscillations studied in this work, and that the spectra are typically peaked in the $m$ = 2-4 range.  The azimuthal mode number spectrum has important consequences for the interaction of ULF waves with radiation belt electrons \citep[e.g.][]{elkington:99a,hudson:00a,elkington:03a}  and this interaction will be the subject of future work.

\begin{table}
\begin{tabular}{|l||c|c|c|c|}	\hline
& \multicolumn{2}{|c|}{ {\bf Field line A} } & \multicolumn{2}{|c|}{ {\bf Field line B} } \\ \hline
{\bf Run } & Cartesian & Polar & Cartesian & Polar   \\ \hline
10 mHz & (5.0, 5.0, 0) & (7.1, 45$^\circ$, 0) & - & - \\ \hline
15 mHz & (4.6, 4.6, 0) & (6.5, 45$^\circ$, 0) & (6.5, 6.5, 0)  & (9.2, 45$^\circ$, 0) \\ \hline
25 mHz & (4.1, 4.1, 0) & (5.8, 45$^\circ$, 0) & (5.9, 5.9, 0)  & (8.3, 45$^\circ$, 0) \\ \hline 
\end{tabular}
\caption{Locations of the field lines used in the FLR mode structure calculations for the three monochromatic simulations.  The point where the field line intersects the equatorial plane is given in Cartesian, ($x,y,z$), and cylindrical polar, ($r,\varphi,z$), coordinates.  All units are in $R_E$ except $\varphi$, which is in degrees.}
\label{tbl:fllocs} 
\end{table}

%
%
%
%
%
%

%
%
%
%

\begin{acknowledgments}
One author (SGC) wishes to thank his thesis advisors, Scot R. Elkington and Mike Wiltberger, for their outstanding guidance and tutelage.  This material is based upon work supported by the National Aeronautics and Space Administration New Hamsphire Space Grant NNG05GG76H, NASA Grant Nos. NNX08AM34G, NNX08AI36G and by the Center for Integrated Space Weather Modeling, which is funded by the Science and Technology Centers program of the National Science Foundation under agreement ATM-0120950. 
\end{acknowledgments}


%
%
%
%
%
%
%
%
%
%


%
%

\end{article}




%
%
%
%
%
%


\end{document}